# Quantum spin liquid in frustrated one–dimensional LiCuSbO$_4$

S. E. Dutton[1*], M. Kumar[1], M. Mourigal[2,3], Z. G. Soos[1], J.–J. Wen[2], C. L. Broholm[2], N. H. Andersen[4], Q. Huang[5], M. Zbiri[3], R. Toft–Petersen[4] and R. J. Cava[1,4]

[1]Department of Chemistry, Princeton University, Princeton, NJ 08544, USA

[2]Institute for Quantum Matter and Department of Physics and Astronomy, The Johns Hopkins University, Baltimore, MD 21218, USA

[3]Institut Laue–Langevin, BP 156, 38042 Grenoble Cedex 9, France

[4]Department of Physics, Risø Campus, Technical University of Denmark, Building 228, Frederiksborgvej 399, DK-4000 Roskilde, Denmark

[5]Center for Neutron Research, NIST, Gaithersburg, MD 20899, USA

[*]corresponding author: sed33@cam.ac.uk

Abstract

A quantum magnet, LiCuSbO$_4$, with chains of edge–sharing spin–½ CuO$_6$ octahedra is reported. While short–range order is observed for $T < 10$ K, no zero–field phase transition or spin freezing occurs down to 100 mK. Specific heat indicates a distinct high field phase near the 12 T saturation field. Neutron scattering shows incommensurate spin correlations with $q = (0.47\pm0.01)\pi/a$ and places an upper limit of 70 μeV on any spin gap. Exact diagonalization of 16–spin easy–plane spin–½ chains with competing ferro– and antiferromagnetic interactions ($J_1 = -75$ K, $J_2 = 34$ K) accounts for the $T > 2$ K data.

The Heisenberg spin–½ chain is one of very few quantum critical systems to be realized in a crystalline solid. An element of frustration is added by next–nearest–neighbor (NNN) interactions ($J_2$).[1-3] In such systems, theoretical work[4-9] indicates that qualitatively different quantum phases are possible as a function of $\alpha = J_2/J_1$, axial exchange anisotropy, $\Delta$, and the applied field $h = g\mu_B H/|J_1|$. Finite values of $\alpha$ are observed in copper oxide spin–chains formed by corner– or edge–sharing Jahn–Teller distorted $CuO_6$ polyhedra.[10-17] While the sign and magnitude of $J_1$ is dependent on the <Cu–O–Cu bond angles, $J_2$ is always antiferromagnetic (AFM), ensuring frustration. For FM–AFM chains with ferromagnetic (FM) near–neighbor (NN) interactions ($J_1 < 0$), theory indicates a quantum critical point (QCP) at $\alpha_c = -0.25$, $\Delta = 1$,[18,19] which separates a gapless FM state ($\alpha > \alpha_c$) from a short–range ordered (SRO) phase with a small yet finite gap ($\alpha < \alpha_c$).[20] Dimer, spiral, vector–chiral and multipolar phases have been predicted as a function of $\alpha$, $\Delta$ and $h$.[4,6,8] Interest has been further amplified by the observation that ferroelectricity accompanies magnetic ordering in some of these systems.[14,21,22]

Known spin–½ FM–AFM chains span a wide range of $\alpha$ and include $Li_2CuO_2$,[17,21,22] $SrCuO_2$,[15] $LiCuVO_4$,[11,14,23] $Li_2CuZrO_4$[10] and $A_2Cu_2Mo_3O_{12}$ (A = Rb,Cs).[12,13] With the exception of $A_2Cu_2Mo_3O_{12}$ (A = Rb,Cs), all order in three–dimensions (3D) and cannot be fully magnetised in $Nb_3Sn$ superconducting electromagnets; the exception, has a more complex structure where Dzyaloshinskii–Moriya interactions dominate the low $T$ behaviour.[24]

Here we present the structure and magnetic properties of the one–dimensional (1D) spin chain compound $LiCuSbO_4$. Due to weak inter–chain interactions it has no 3D ordering down to 100 mK and can be driven to a saturated–FM phase above the experimentally accessible field $<\mu_o H_s> = 12$ T. The magnetic susceptibility, isothermal magnetization and specific heat can be simultaneously modelled for $T > 2$ K using the approximate spin Hamiltonian:

$$H = J_1 \sum_r \left( \frac{3}{2+\Delta}(S_r^x S_{r+1}^x + S_r^y S_{r+1}^y + \Delta S_r^z S_{r+1}^z) + \alpha S_r \cdot S_{r+2} - h(S_r^z \cos\theta + S_r^x \sin\theta) \right) \quad (1)$$

Here $\theta$ is the angle between the applied field, $h$, and the unique molecular axis, $z$. The parameters $J_1 = -75$K, $\alpha = -0.45$, and $\Delta = 0.83$ account for the main features in the thermomagnetic data and are also consistent with the magnetic excitations measured by inelastic neutron scattering.

Powder samples of $LiCuSbO_4$ were prepared using a ceramic synthesis.[25] $LiCuSbO_4$ has an orthorhombic structure[26,27] related to that of $LiFeSnO_4$.[28] Rietveld analysis[29] of powder neutron diffraction data acquired on the BT1 diffractometer at NIST[30] was carried out using *FULLPROF*.[31] The final fit, with the corresponding crystal structure, is displayed in Fig. 1, the structure is given in Table 1. Further details of the structural refinement are given in ref. 30. In LiCuSbO$_4$, edge–sharing CuO$_6$ octahedra form chains along the $a$ axis and our analysis limits Li or Sb defect occupancy on the Cu sites to less than 1 %. As expected for $d^9$ Cu$^{2+}$ ions, a Jahn–Teller distortion results in significant elongation of the axial Cu–O bonds. The relatively low symmetry allows for some alternation of Cu–O bond lengths and angles along the chains (Fig. 1(b)), although only a single Cu–Cu intra–chain distance occurs. While the connectivity of the Cu atoms within the chains is similar to that of recently studied frustrated FM–AFM spin–½ chain compounds,[10-15] the easy–axis of the CuO$_6$ octahedra in adjacent chains are not parallel. This unique feature of LiCuSbO$_4$ may be responsible for weaker inter–chain interactions which suppress 3D ordering.

Magnetization measurements on a powder sample of LiCuSbO$_4$ were carried out using a Quantum Design Physical Properties Measurement System (PPMS) with a 9 T magnet and a CRYOGENIC Cryogen Free Measurement System (CFMS) with a 16 T magnet. Magnetic susceptibility ($\chi = dM/dH$) measurements (Fig. 2(a)) were performed after cooling in either zero–field (ZFC) or the measuring field (FC). In lower field measurements, $\mu_o H \leq 11$ T, a broad maximum in $\chi$, characteristic of SRO, is observed at $T = T_m$. In applied fields $T_m$ initially increases from ~6 K at 0.1 T to a maximum ~9 K at 2 T. In higher field a decrease in $T_m$ is observed (Fig. 3(c)) until at 16 T, $\chi$ continues to increase monotonically upon cooling indicating a crossover to field induced FM. The absence of a low $T$ Curie-tail in the SRO order regime indicates the high quality of our sample and confirms the absence of any Cu site disorder. As has been noted elsewhere,[32-35] fitting $\chi^{-1}$ of FM-AFM chains to the Curie-Weiss law is highly dependent on the fitting criteria. To reduce the uncertainty in our analysis the magnetic moment of the Cu$^{2+}$ cations was fixed at $\mu_{eff} = g_{av}\sqrt{S(S+1)}$. The average g–factor, $g_{av} = 2.10$, was determined by electron paramagnetic resonance (EPR) measurements at 4 K using a Bruker K–band spectrometer and allows us to experimentally determine $\mu_{eff} = 1.82$ $\mu_B$ per fu. A representative fit of the high $T$ $\chi^{-1}$ to the Curie-Weiss law, $\chi - \chi_0 = C/(T-\theta)$, for $420 < T < 600$ K is inset in Fig. 2(a). Whilst not unique due to the ambiguity in defining $\chi_0$, the value obtained $\theta = 25(1)$ K is as would be expected for dominant FM correlations.

Isothermal magnetization measurements up to 16 T indicate proximity to a FM phase (Fig. 2(b)). For $T < T_m$, the magnetization has an 'S' shape curvature associated with an increase in the uniform FM component. At 2 K the magnetization approaches saturation in the limiting 16 T field. $dM/d(\mu_o H)$ shows a maximum when $T < T_m$ indicating an orientation averaged saturation field, $\mu_o\langle H_s\rangle$ = 12 T, in the low temperature limit. Our measurements clearly indicate that 3D ordering is absent in LiCuSbO$_4$ down to 0.1 K in zero–field. The low saturation field, $\mu_o\langle H_s\rangle$ = 12 T, is a further indication of very weak inter–chain interactions.[36]

Specific heat, $C_p$, measurements were preformed using a Quantum Design PPMS. To improve thermal conductivity, pellets of LiCuSbO$_4$ mixed with silver were used for our measurements. The $C_p$ with the contribution from the silver powder[37] deducted is shown in Fig. 3(a)–(c). In 0 T a broad maximum in $C_p$ characteristic of the onset of SRO is observed for $T \sim 7.0$ K, this corresponds to a maximum in $C_p/T$ at 4.3 K. With increasing field the peak shifts to lower temperature, until at 10 T it is completely suppressed (Fig. 3(b)–(c)). The lattice contribution to the $C_p$ was modeled using the Debye Law ($\theta_D$ = 410 K), and the magnetic entropy, $\Delta S = \int C_p/T \, dT$, was extracted ($T$ > 0.1 K) (Fig. 3(a) inset). At high temperatures $\Delta S$ measured in zero and an applied field converges and at 50 K the 0 T measurement approaches the expected value ($R\ln 2$). At lower temperatures, $T$ < 2 K, a second anomaly, $T \sim 0.6$ K, enhanced on application of a field, is observed for $\mu_o H \leq 13$ T. One possible origin for this anomaly is a multipolar transition as predicted theoretically for $H \sim H_s$.[4-6, 8, 9] However, interpretation of high field measurements in LiCuSbO$_4$ are complicated not only by the spherical averaging inherent to a powder sample but also because the alternating tilts of the easy–axis of the CuO$_6$ octahedra results in an effective staggered field for all field orientation except $H \parallel a$. This can open a gap and preclude a phase transition.[38] Indeed, the broad nature of the peak in the $C_p/T$ indicates a crossover rather than a phase transition. Single crystal experiments and more detailed theoretical work are needed to establish the nature of the high field phase suggested by the powder measurements.

To determine the relevant spin Hamiltonian, exact diagonalization (ED) of up to 20–spin rings was used to simultaneously model the magnetic susceptibility, isothermal magnetization and specific heat.[39] Almost the complete temperature (2 – 300 K) and field (0 – 16 T) range of our thermomagnetic measurements was used to refine the ED model. As a generic first approximation to FM–AFM spin–½ chains,[4, 12, 40] the model neglects g–tensor anisotropy and dipolar, hyperfine and inter–chain interactions between spins. To reduce the number of parameters, an average g–factor, $g_{av}$ = 2.10, determined by EPR was used. Initially the magnetic properties were modeled with an isotropic chain, $\Delta$ = 1. With $\alpha$ = –0.45 and $J_1$ = –68 K a good description of the thermodynamic data is obtained, though the high field magnetization is overestimated (Fig. 2(b)). To account for this we included easy–plane anisotropy for the NN interactions, for simplicity leaving the NNN interactions isotropic. We also neglect the more subtle effects of small deviations from an ideal square planar geometry. The model with $J_1 = -75$ K, $\alpha = -0.45$, and $\Delta = 0.83$ for 16–spins correctly reproduces the isothermal magnetization at 2 K (Fig. 2(b)). This dramatic improvement in the fit to the isothermal magnetization when $\Delta = 0.83$ is due to a range of critical fields for a spherically averaged anisotropic model. $\Delta$ is of similar magnitude to that expected for Jahn–Teller distorted $Cu^{2+}$, $\Delta \sim 0.9$.[4] The presence of two distinct Cu–Cu exchange pathways allows for a more complex $J_1$, $J'_1$, $J_2$ exchange model, a 10% modulation of $J_1$ has no significant effect on the predicted properties. Given the insensitivity of the model to variation in $J_1$ we have not considered additional exchange interactions.

To explore the nature of the magnetic excitations at 0 T and test the magnetic exchange model inferred from the thermomagnetic measurements, inelastic neutron scattering on a powder sample of $^7$LiCuSbO$_4$ was carried out using the IN6 Cold neutron time–of–flight time–focusing spectrometer at the Institute Laue Langevin (ILL). The powder-averaged dynamical structure factor, $S(Q,\omega)$, measured at $T = 1.5$ K and $T = 6.0$ K (Fig. 4(a)–(b)) displays intensity at low momentum–transfer which can be unambiguously associated with magnetic excitations. For $T = 1.5$ K the spectrum appears to be gapless within the 70 μeV elastic energy resolution of the instrument and most of the intensity is observed below 1.8 meV; no higher energy intensity is detected below the maximal energy transfer accessible with $E_i = 3.12$ meV ($\lambda = 5.12$ Å). Comparing the 1.5 K and 6 K spectra reveals that a magnetic quasi–elastic signal, $\hbar\omega < 0.30$ meV, develops at low–temperature. As a function of momentum–transfer this signal is most intense for $Q = 0.52(1)$ Å$^{-1}$ (Fig. 4(d)). This corresponds to $Q = 0.475(9)$ π/$a$, and is consistent with short–ranged incommensurate correlations in a frustrated spin–chain with $\alpha = -0.45$.[40] Remarkably given the exchange parameters extracted from thermodynamic data, half of the magnetic scattering is found for $0.7 \leq \hbar\omega \leq 2.0$ meV (Fig. 4(b) and 4(e)). The spectral maximum, $\hbar\omega \sim 0.7$ meV, is consistent with the 4.3 K maximum in the zero–field specific heat. To rigorously test the simplified spin Hamiltonian, we calculated the spherically averaged[41, 42] dynamic response function, $S(Q,\omega)$, with the parameters (N = 16, $J_1 = -75$ K, $\alpha = -0.45$ $\Delta = 0.83$) extracted from the thermodynamic data. The result (Fig. 4(e)) is remarkably consistent with the measured data given the absence of adjustable parameters.[43]

LiCuSbO$_4$ is a novel crystalline quantum–magnet with no 3D zero–field phase transition down to $T = 100$ mK. Using a combination of structural refinement, bulk magnetic properties, ED, and inelastic neutron scattering we show that these unusual properties result from spin–½ chains with frustrated FM–AFM interactions ($\alpha = -0.45$). The extremely low and experimentally accessible saturation field of 12 T makes LiCuSbO$_4$ an ideal material in which to explore, in its entirety, the rich 1D physics predicted for frustrated spin–½ chains near quantum criticality.

Acknowledgements

The authors wish to acknowledge the assistance of C. Pacheco during the EPR measurements. RJC acknowledges support from the Velux Visiting Professor Programme 2009–2010 during his visit to Risø DTU. The ILL facility is acknowledged for providing beamtime on the IN6 spectrometer. This research was supported by the U.S. Department of Energy, Office of Basic Energy Sciences, Division of Materials Sciences and Engineering under Award DE–FG02–08ER46544.

Table 1: Structural parameters for LiCuSbO$_4$ obtained from neutron powder diffraction at 298 K. See ref. 30 for further details. $^{\pm}$LiSbO$_3$ is nonmagnetic and so does not effect our thermomagnetic measurements. $^{*}$The fraction of CuO ($T_N$ = 230 K) in our sample is too small to be detected in our measurements. At low $T$ all of the Cu$^{2+}$ spins in CuO are frozen in an AFM ordered state, the absence of a spin tail in the magnetic susceptibility (Fig. 2(a)) confirms this.

| **LiCuSbO$_4$** | **298 K** | **C$mc2_1$ (36)** | **Z = 8** | | |
|---|---|---|---|---|---|
| $a$ = 5.74260(4) $b$ = 10.86925(7) Å $c$ = 9.73048(6) Å $V$ = 609.528(7) Å$^3$ | | | | | |
| $R_{wp}$ = 8.90 | | | | | |
| | $x$ | $y$ | $z$ | $B_{iso}$ / Å$^2$ | Frac |
| Sb1 (4$a$) | 0 | 0.3108(3) | 0.2696(4) | 0.13(5) | 1.0 |
| Sb2 (4$b$) | 0 | 0.1677(4) | −0.0050(5) | 0.16(5) | 1.0 |
| Cu (8$b$) | 0.7500(0) | 0.41030(15) | 0.0(0) | 0.49(3) | 1.0 |
| Li1a (4$a$) | 0 | 0.344(2) | 0.6974(13) | 0.3(2) | 0.5 |
| Li1b (4$a$) | 0 | 0.378(2) | 0.6974(13) | 0.3(2) | 0.5 |
| Li2 (8$b$) | 0.026(4) | 0.0002(12) | 0.2660(13) | 0.5(3) | 0.5 |
| O1 (4$a$) | 0 | 0.0039(3) | 0.0788(4) | 0.45(6) | 1.0 |
| O2 (4$a$) | 0 | 0.3264(4) | 0.9026(4) | 0.37(7) | 1.0 |
| O3 (4$a$) | 0 | 0.4538(3) | 0.1400(5) | 0.43(5) | 1.0 |
| O4 (8$b$) | 0.2265(4) | 0.2297(2) | 0.1360(3) | 0.36(3) | 1.0 |
| O5 (4$a$) | 0 | 0.1568(4) | 0.3788(6) | 0.57(6) | 1.0 |
| O6 (8$b$) | 0.7447(7) | 0.3957(2) | 0.3704(5) | 0.38(4) | 1.0 |
| LiSbO$_3$$^{\pm}$ 2.73(10) wt% | CuO$^{*}$ 0.43(5) wt% | | | | |

Fig. 1 (color online): (a) Observed (o) and calculated (–) neutron diffraction data for LiCuSbO$_4$ collected at 298 K. The difference curve is also shown; reflection positions are indicated by the vertical lines for LiCuSbO$_4$ (upper), and the LiSbO$_3$ (2.73(10)) wt%) (middle) and CuO (0.43(5) wt%) (lower) impurity phases. A polyhedral model looking down the CuO$_6$ chains is inset. The Jahn–Teller distorted CuO$_6$ polyhedra are represented with the long axial Cu–O bonds omitted. (b) Ball and stick model showing the connectivity of the Cu chains in LiCuSbO$_4$. Cu–Cu and Cu–O bond lengths are black and blue respectively, Cu–O–Cu bond angles are red.

Fig. 2 (color online): (a) ZFC and FC magnetic susceptibility of LiCuSbO$_4$ in a 0.1 T and 16 T measuring field. The calculated susceptibilities for a 16–spin anisotropic frustrated chain model $\alpha = -0.45$, $J_1 = -75$ K, $\Delta = 0.83$, $g_{av} = 2.10$ are shown in a 0.1 T (dashed) and 16 T (solid) field. The high temperature inverse magnetic susceptibility in 8 T and the fit to the Curie-Weiss law ($\chi_0 = 7.36 \; 10^{-5}$ emu Oe$^{-1}$ mol$^{-1}$) is inset. (b) Isothermal magnetization of LiCuSbO$_4$ and corresponding predictions for the frustrated chain model. At 2 K an isotropic (brown) 24–spin system, $\alpha = -0.45$, $J_1 = -68$ K is shown in addition to the anisotropic (black) model. dM/dµ$_o$H is plotted below, for $T < T_m$ a clear maximum at µ$_o$H ~ 12 T is observed.

Fig. 3 (color online): (a) $C_p$ as a function of field and temperature for selected fields. The lines indicate fits to the 16–spin anisotropic frustrated chain model, $\alpha = -0.45$, $J_1 = -75$ K, $\Delta = 0.83$, $g_{av} = 2.10$ in a 0 T (black), 9 T (red) and 16 T (green) applied field. The magnetic entropy, $\Delta S = \int C_p/T \; dT$ integrated from 0.1 K is inset. (b) $C_p/T$ as a function temperature plotted on a logarithmic scale. (c) Contour plot of $C_p/T$ as a function of field and temperature for LiCuSbO$_4$. The SRO transition as indicated by magnetic susceptibility (squares) is presented as a function of magnetic field and temperature.

Fig. 4 (color online): (a) Low temperature inelastic neutron scattering data at $T = 1.5$ K and (b) $T = 6.0$ K. (c) Calculation of $S(Q,\omega)$ in absolute units for a spin–16 chain at $T = 0$ K using ED with $\alpha = -0.45$, $J_1 = -75$ K, $\Delta = 0.83$. (d) Momentum-transfer dependence of the higher-energy signal, $0.50 < \hbar\omega < 1.75$ meV and (e) quasielastic signal, $0.18 < \hbar\omega < 0.31$ meV. The color scale is in absolute units and consistent for frames (a)-(c). The intensity of the red solid line in (e) was rescaled by a factor of 1/2 to compare with experimental results.

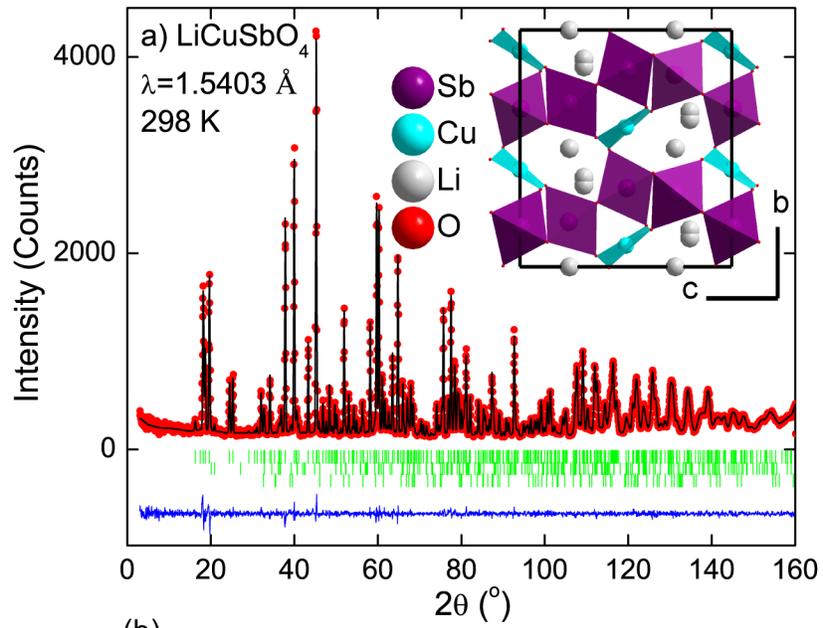

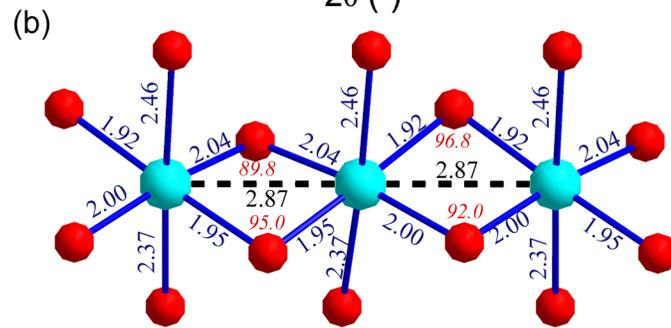

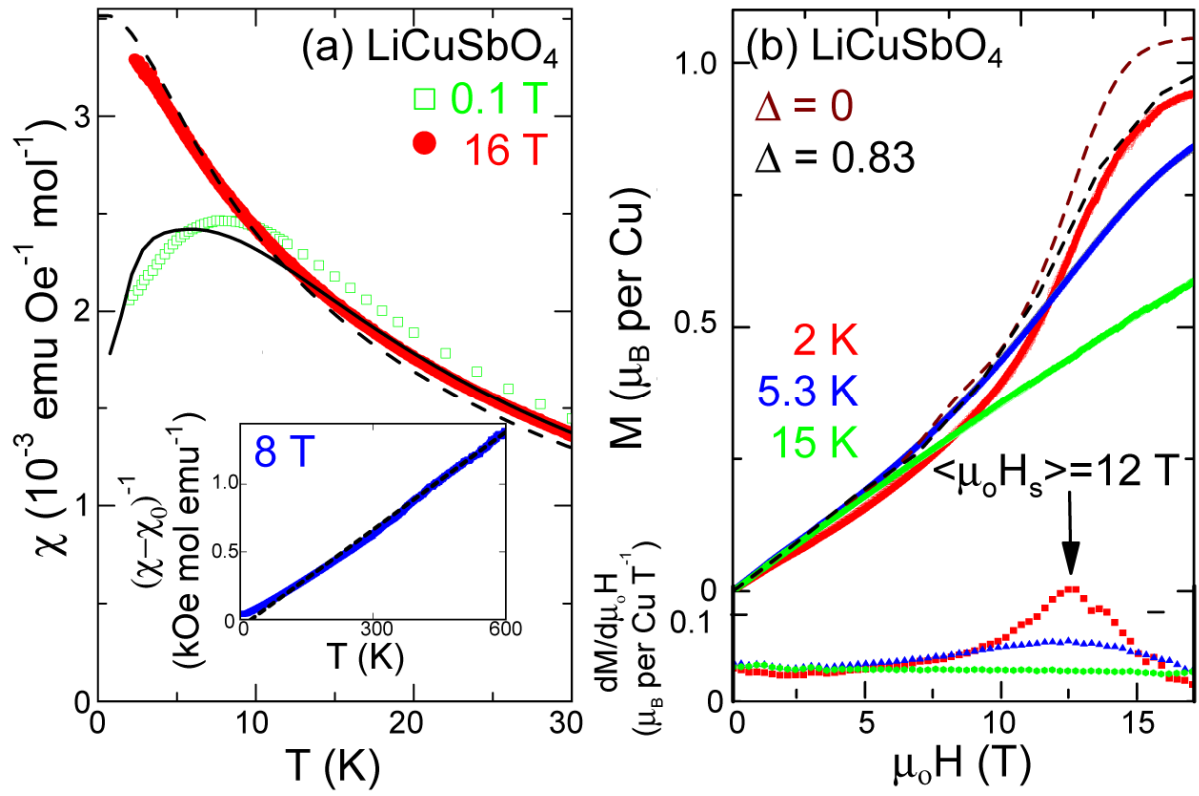

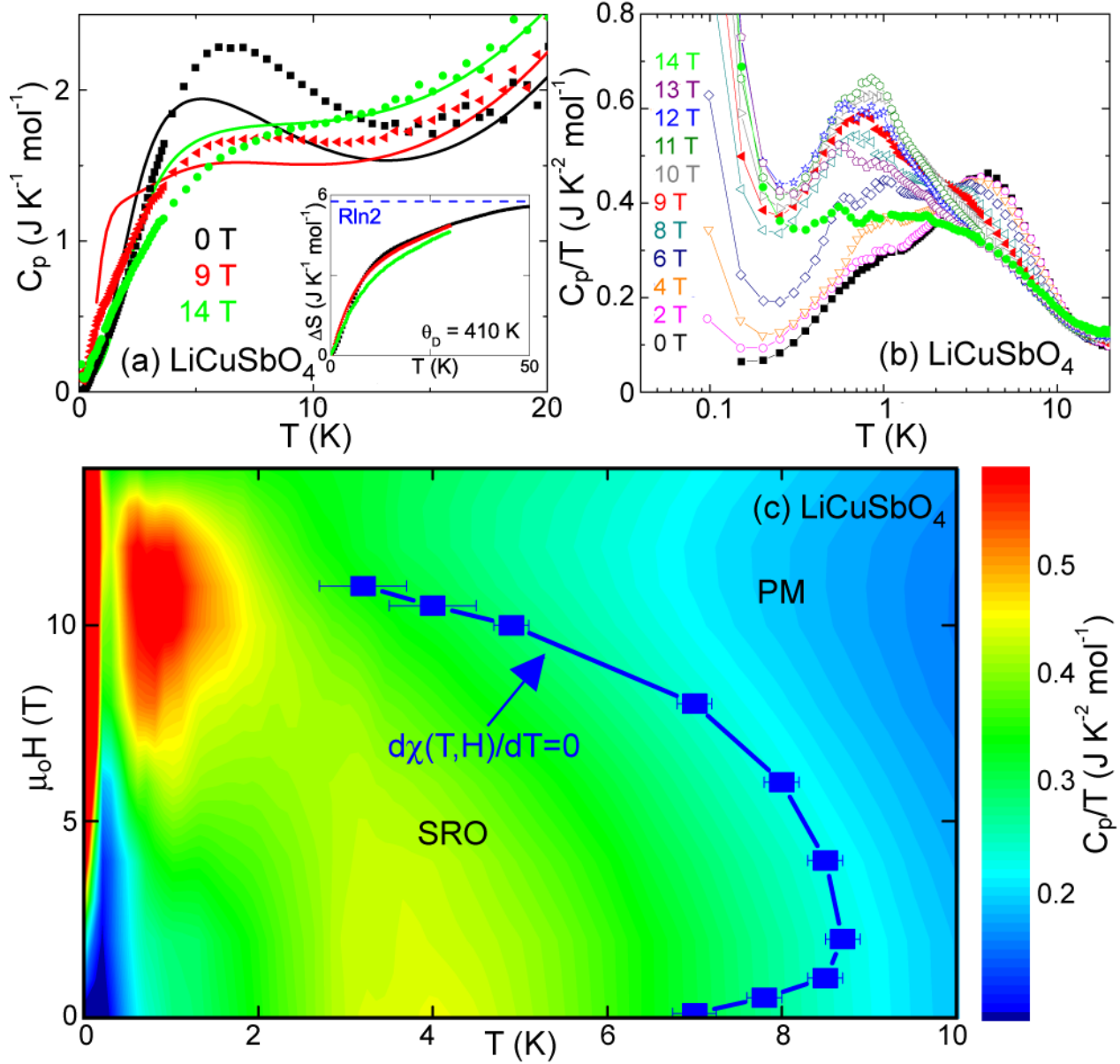

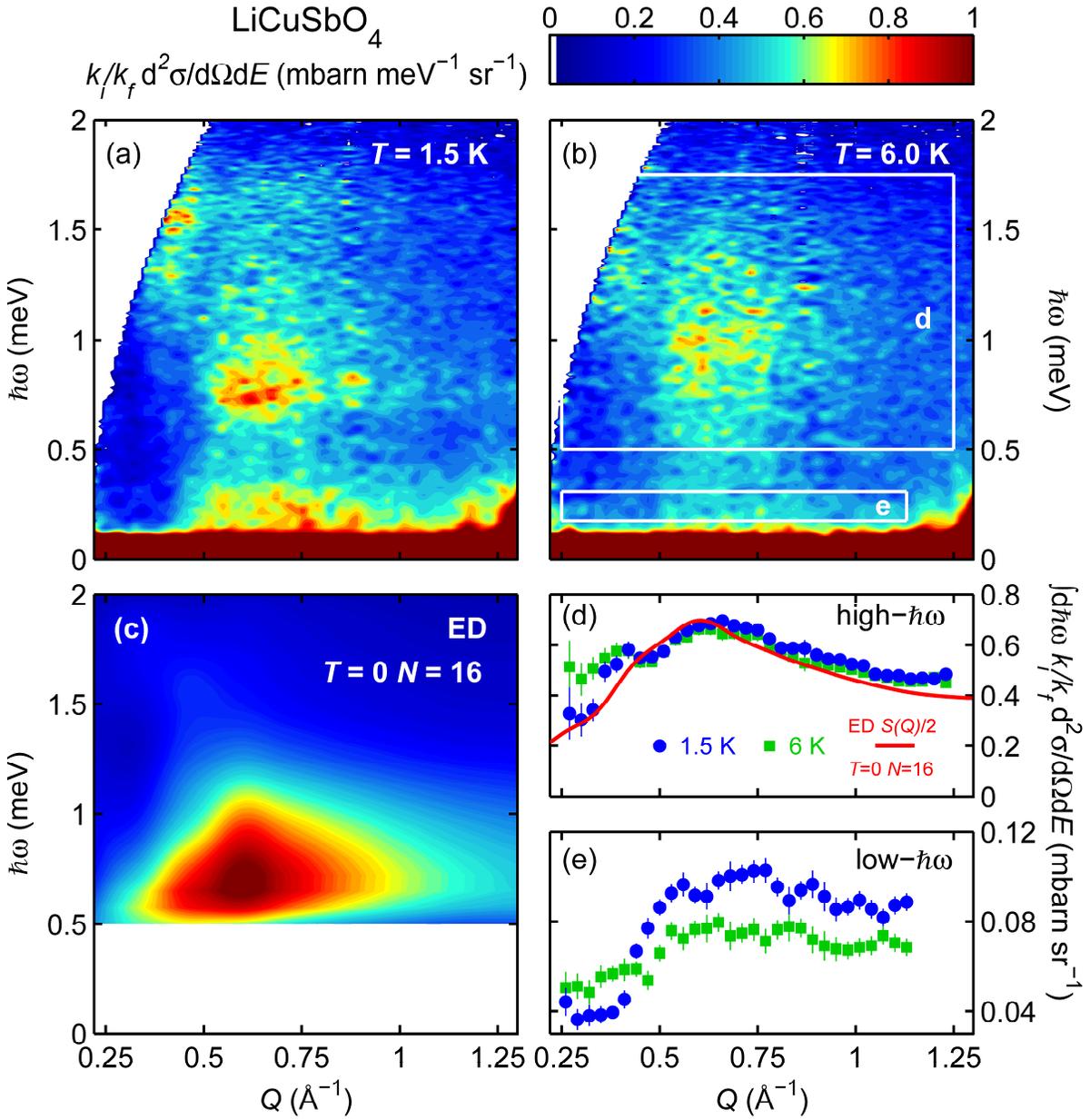

Fig. 4